\documentclass{iaus}
\usepackage{graphicx}

\title[Probing ISM Magnetic Fields With SNRs] 
{Probing Interstellar Magnetic Fields With Supernova Remnants}

\author[Kothes \& Brown]   
{Roland Kothes$^1$ \and Jo-Anne Brown$^2$}

\affiliation{$^1$National Research Council of Canada,
              Herzberg Institute of Astrophysics,
              Dominion Radio Astrophysical Observatory,
              P.O. Box 248, Penticton, British Columbia, V2A 6J9, Canada 
	      \break email: Roland.Kothes@nrc-cnrc.gc.ca\\[\affilskip]
$^2$ Department of Physics and Astronomy, University of Calgary,
             2500 University Drive N.W., Calgary, AB, Canada 
	     \break email: jocat@ras.ucalgary.ca}

\pubyear{2009}
\volume{259}  
\pagerange{100--100}
\date{December 9, 2008}
\jname{Cosmic Magnetic Fields: From Planets, to Stars
and Galaxies}
\editors{K.G. Strassmeier, A.G. Kosovichev \&
J.E. Beckman, eds.}
\begin{document}

\maketitle

\begin{abstract}
As supernova remnants expand, their shock waves are freezing in and compressing 
the magnetic field lines they encounter; consequently we can use supernova 
remnants as magnifying glasses for their ambient magnetic fields. We will 
describe a simple model to determine emission, polarization, and rotation 
measure characteristics of adiabatically expanding supernova remnants and how 
we can exploit this model to gain information about the large scale magnetic 
field in our Galaxy. We will give two examples: The SNR DA530, which is located 
high above the Galactic plane, reveals information about the magnetic field in 
the halo of our Galaxy. The SNR G182.4+4.3 is located close to the anti-centre 
of our Galaxy and reveals the most probable direction where the large-scale 
magnetic field is perpendicular to the line of sight. This may help to decide 
on the large-scale magnetic field configuration of our Galaxy.
\end{abstract}

\firstsection 
\section{Introduction}

Recently, there have been several studies of the Milky Way's
magnetic field by the means of observing the rotation measure of compact 
polarized objects like extra-galactic point sources (e.g. \cite{brown07}) or 
pulsars (see talks by A. Noutsos and J.-L. Han). However, we 
still do not know the large-scale magnetic field configuration or even the
number of field reversals within our Galaxy. 
One problem is that through 
Faraday rotation studies of extra-galactic point sources we only derive the
average magnetic field parallel to the line of sight $B_\parallel$ through our 
Galaxy weighted by the electron density $n_e$, because the rotation measure 
$RM$ is given by:
\begin{equation}
RM = 0.81 \int_l B_\parallel n_e dl,
\label{RM}
\end{equation}
here $RM$ is given in rad/m$^2$, $B$ in $\mu$G, $n_e$ in cm$^{-3}$, and the
pathlength $l$ in pc. In addition extragalactic sources may suffer from 
intrinsic Faraday rotation of unknown magnitude. Faraday rotation studies of
pulsars average $B_\parallel$
between us and the pulsar weighted by $n_e$. In addition pulsars distances
are usually quite uncertain and pulsars probe only a very small area in space
since they are only about 15\,km in diameter. One big problem of the 
averaging
procedure is that there could be numerous field reversals along the line of 
sight, which
would be averaged out. This ambiguity could be solved if we had anchor 
points for the magnetic field within our Galaxy. We propose to determine these 
anchor points
with polarization and
Faraday rotation studies of supernova remnants (SNRs), since these
can be used as magnifying glasses of their ambient magnetic field.

\section{A simple model of supernova remnants}
\begin{figure}
\begin{center}
  \includegraphics[bb = 60 75 650 450,width=7cm,clip]{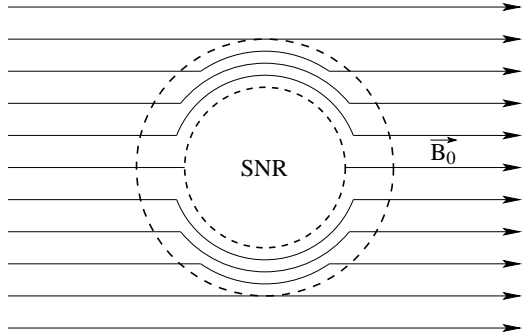}
\end{center}
\caption{Simple model of the magnetic field inside an
adiabatically expanding SNR.}
\label{sketch}
\end{figure}

We assume that the supernova remnants are adiabatically expanding into an 
ambient medium of constant density $n_0$ and constant homogenous magnetic field
$\vec{B}_0$. In our model the SNR is circular and has a shell with a width of 
about 10\,\% of the SNR's radius in which the ambient medium and 
its magnetic field are compressed by a factor of 4. While the electron density in
the shell is assumed to be constant the magnetic field changes with the angle 
between the SNR's expansion direction and the ambient magnetic field 
(see Fig.~\ref{sketch}). It is proportional to the ``number'' of magnetic field lines, which
are swept up by the expanding shock wave, hence the magnetic field inside the 
shell is strongest where the SNR is expanding perpendicular to the ambient 
magnetic field and 0, where it is parallel.
The synchrotron emission is integrated along the line of sight from the back of 
the SNR to the front and appropriately Faraday rotated (see Equation~\ref{RM}). 
The synchrotron flux density $S$ produced by a spectrum of 
relativistic electrons $N(E)$ at frequency $\nu$ is given by:
\begin{equation}
S_\nu \propto K B_\perp^{\frac{1}{2} (\delta + 1)} \nu^{-\frac{1}{2} 
(\delta + 1)},~~~N(E)dE = K E^{-\delta} dE,
\label{flux}
\end{equation}
here $E$ is the energy and the values for $K$ and $\delta$ are defined by 
$N(E)$.

\begin{figure}
\begin{center}
  \includegraphics[width=13cm,clip]{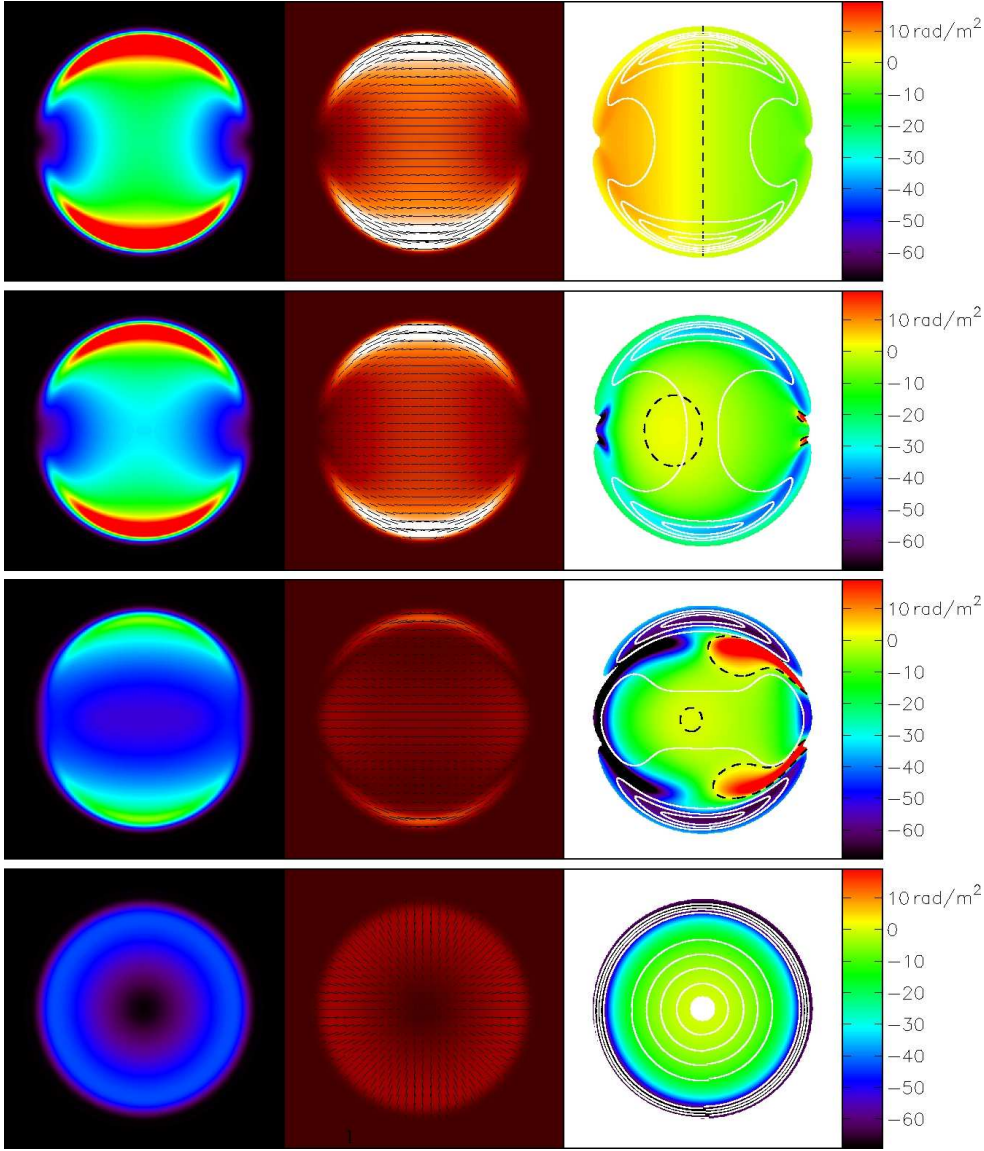}
\end{center}
\caption{Modeled emission structure in Stokes I (left column), polarized 
intensity (centre column) with overlaid B-vectors, and rotation measure 
(right column) with overlaid white contours indicating the polarized emission
for different viewing angles $\Theta$ (from top to bottom: $\Theta = 0^\circ$, 
$30^\circ$, $60^\circ$, and $90^\circ$).}
\label{model}
\end{figure}

\begin{figure}
\begin{center}
\begin{minipage}{5cm}
  \includegraphics[bb = 50 45 530 490,width=5cm,clip]{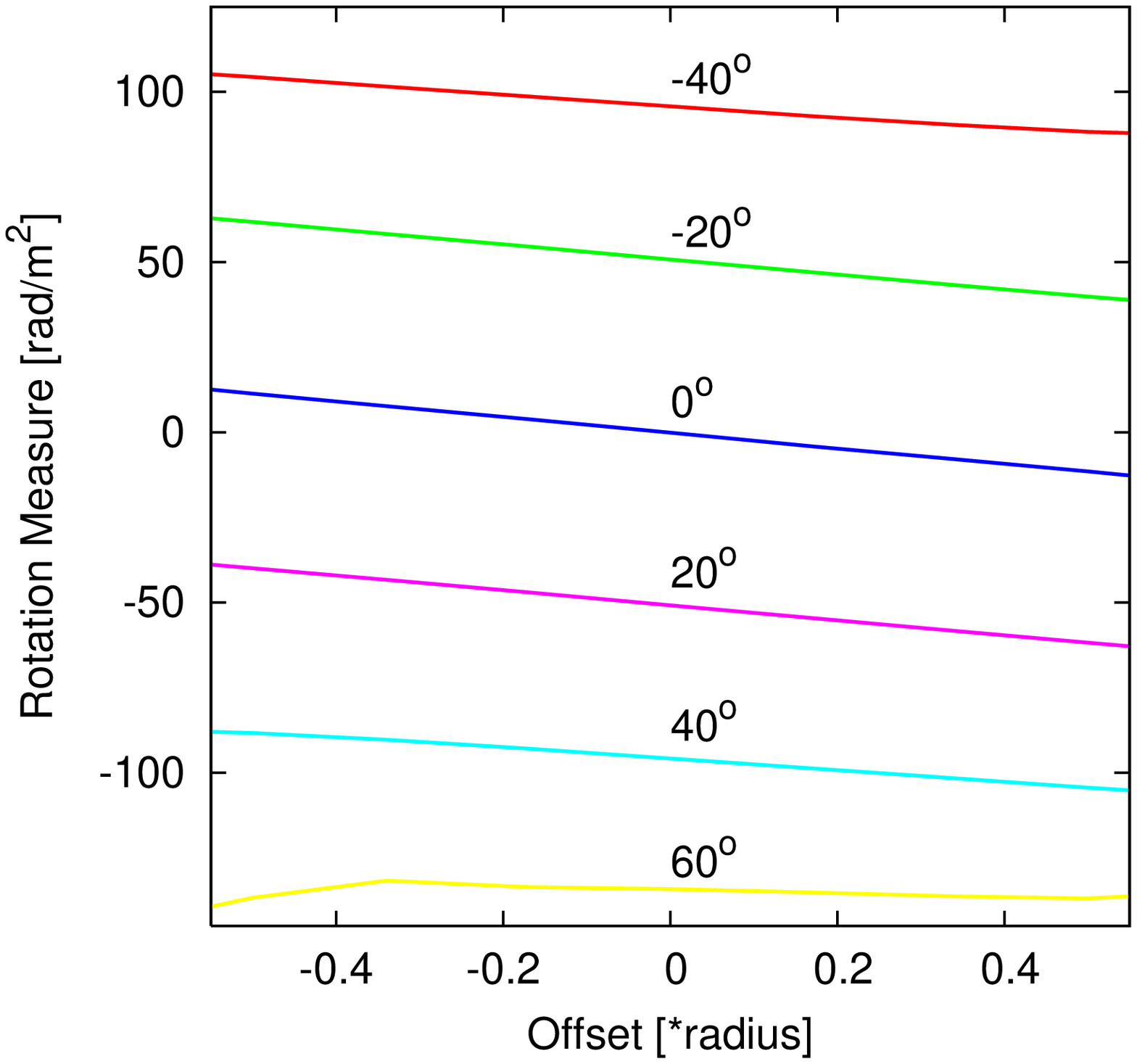}
\end{minipage}
\begin{minipage}{5cm}
  \includegraphics[bb = 50 45 530 490,width=5cm,clip]{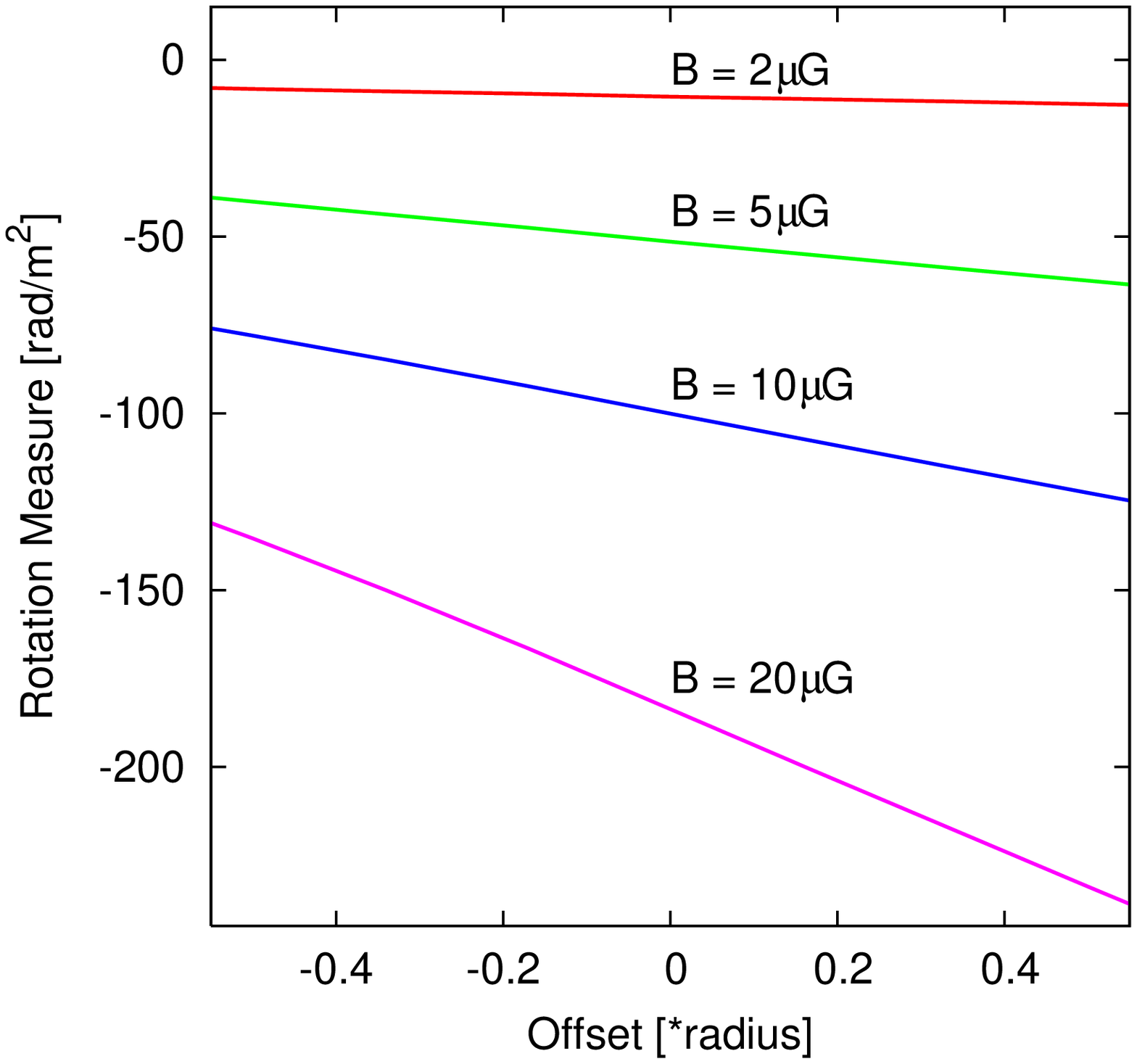}
\end{minipage}
\end{center}
\caption{The rotation measure gradient on the shells of the SNR in the simulation 
described in section 2.2..Left: $\Theta$ varies from $-40^\circ$ (top) to 
$+60^\circ$ (bottom). Right: $\Theta = 20^\circ$ and the ambient magnetic field
varies from $2\mu$G (top) to $20\mu$G (bottom).}
\label{grad}
\end{figure}

In Fig.~\ref{model} we display the simulated emission structure in total 
intensity
and polarized intensity and the internal rotation measure that would be observed
from a typical supernova remnant at different viewing angles $\Theta$. In this
simulation the ambient magnetic field is pointing away from us from the front left to the 
back right and $\Theta$ is the angle between the plane of the sky and the 
magnetic field lines the SNR is expanding into. For negative $\Theta$ here, the 
magnetic field would point towards us from the back left to the front right.
In the simulation shown in Fig.~\ref{model} the ambient magnetic field is $4\,\mu$G, the ambient density
is 1\,cm$^{-3}$, the distance of the SNR is 2\,kpc, and its diameter is 20\,pc.
The simulation was convolved to a resolution of 2.5'.

In emission we find the typical bilateral structure of SNRs in total power and
polarized intensity with a tangential magnetic field up to a $\Theta$ of about 
$60^\circ$, beyond which the SNR becomes circular and thick-shelled with a radial
magnetic field. The surface brightness of the SNR is decreasing form 
$\Theta = 0^\circ$ to $\Theta = 90^\circ$ by a factor of almost 10. The rotation
measure structure reveals a few very interesting characteristics. The internal 
rotation measure in the centre of the SNR is always 0, which indicates that
from observations we can determine the foreground $RM$ there. The entire SNR
except for $\Theta = 0^\circ$ is dominated by RM of one sign; 
in Fig.~\ref{model} it
is mostly negative. This indicates whether the ambient magnetic field is pointing
away from us or towards us. There is also a smaller area of opposite sign. Its 
location indicates the orientation of the ambient magnetic field and its distance 
from the centre gives an approximate value for $\Theta$. In addition we find that
the RM on the two arcs shows a linear behaviour up to about $\Theta = 60^\circ$
(see Fig.~\ref{grad}). This can 
be described by its 
gradient, which depends entirely on $|\vec{B_0}| \times n_0$ and the size of the
SNR and is almost independent of $\Theta$ and the $RM$ in the 
middle, which is determined from $\Theta$ and the foreground $RM$. To show how we 
can exploit these 
characteristics we will give two examples.

\section{The SNR DA\,530 and the Galactic halo}

\begin{figure}
\begin{minipage}{4cm}
  \includegraphics[bb = 120 115 535 540,height=4cm,clip]{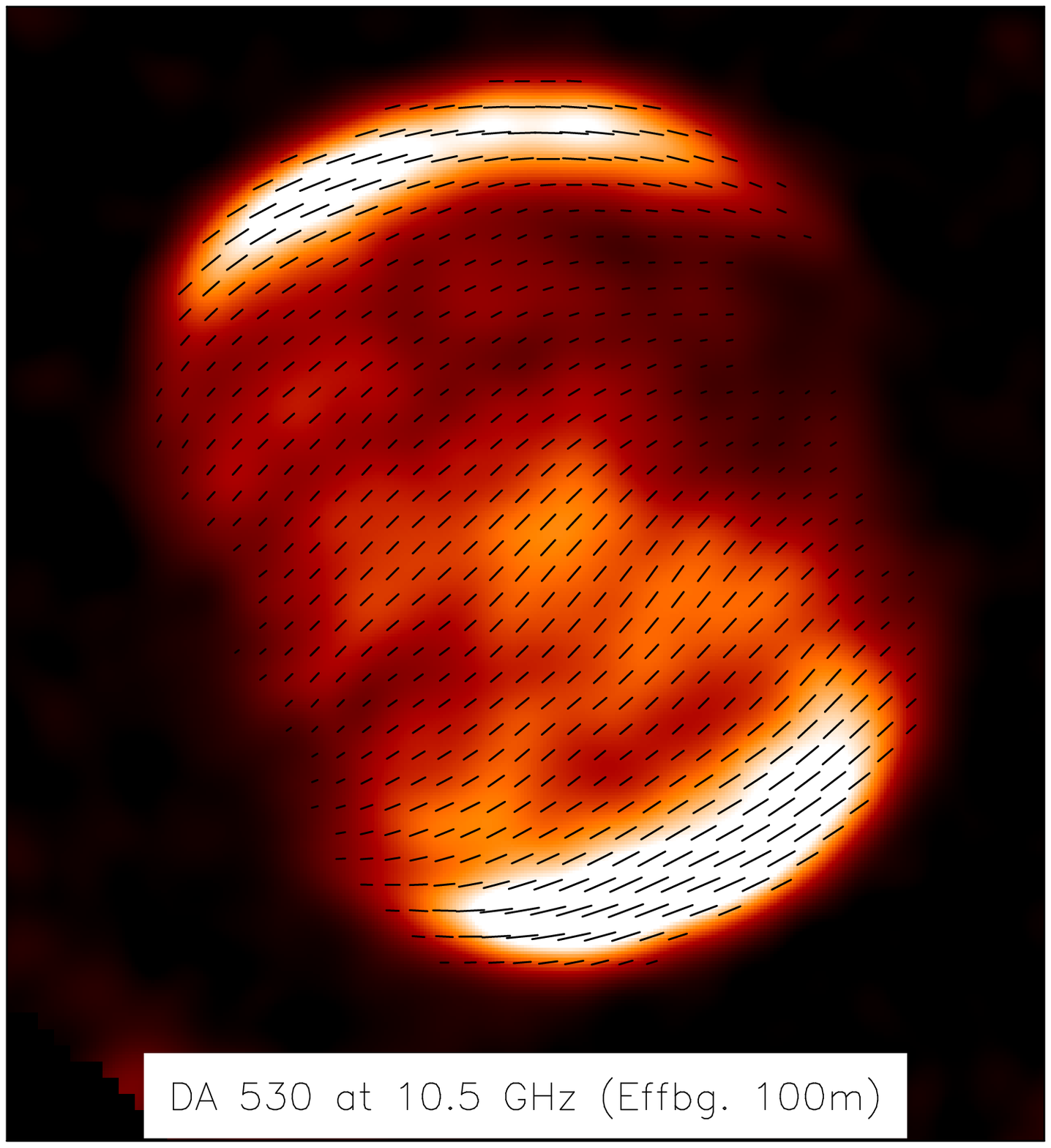}
\end{minipage}
\begin{minipage}{4.8cm}
  \includegraphics[bb = 150 175 545 492,height=4cm,clip]{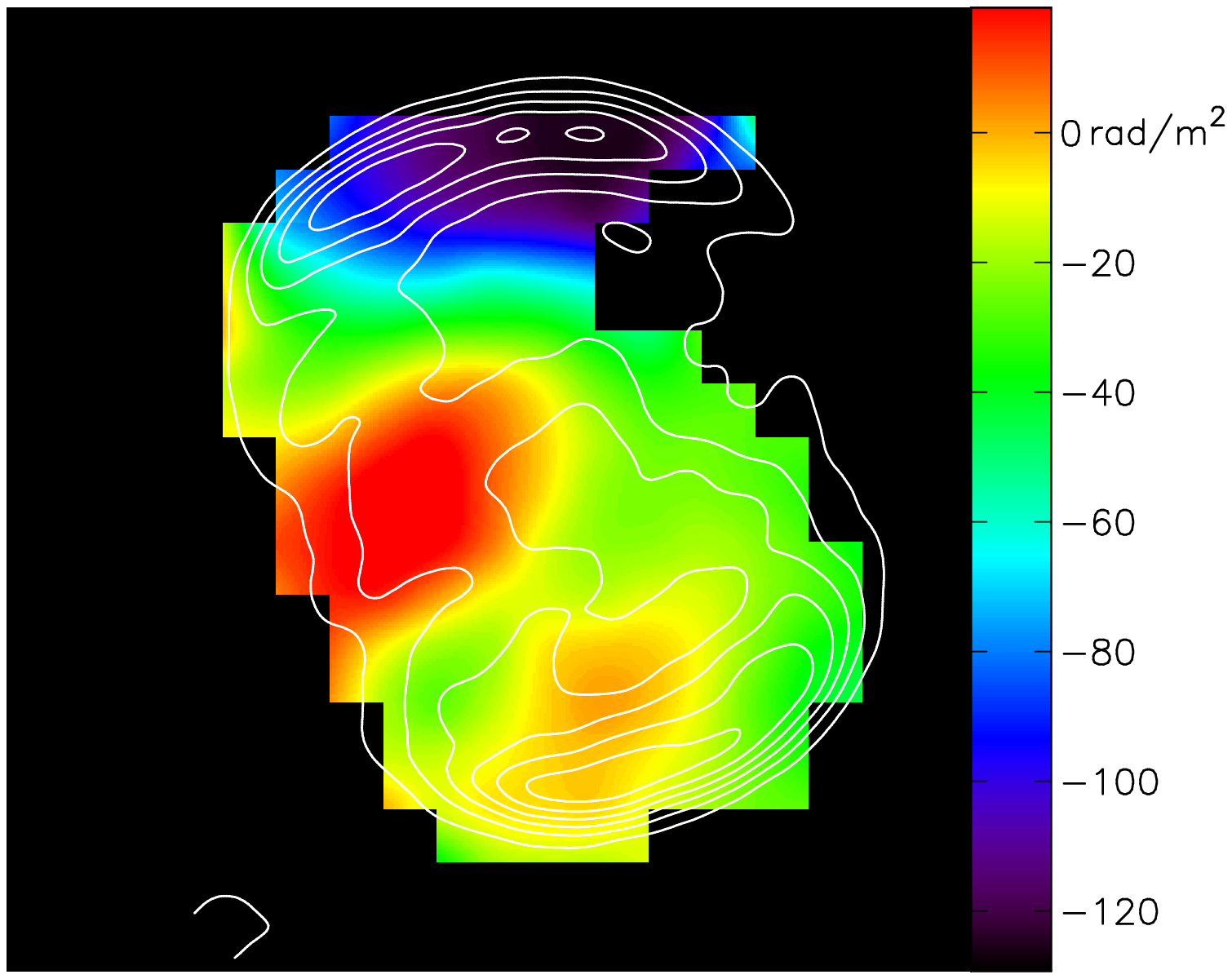}
\end{minipage}
\begin{minipage}{4cm}
  \includegraphics[bb = 65 15 555 465,height=4cm,clip]{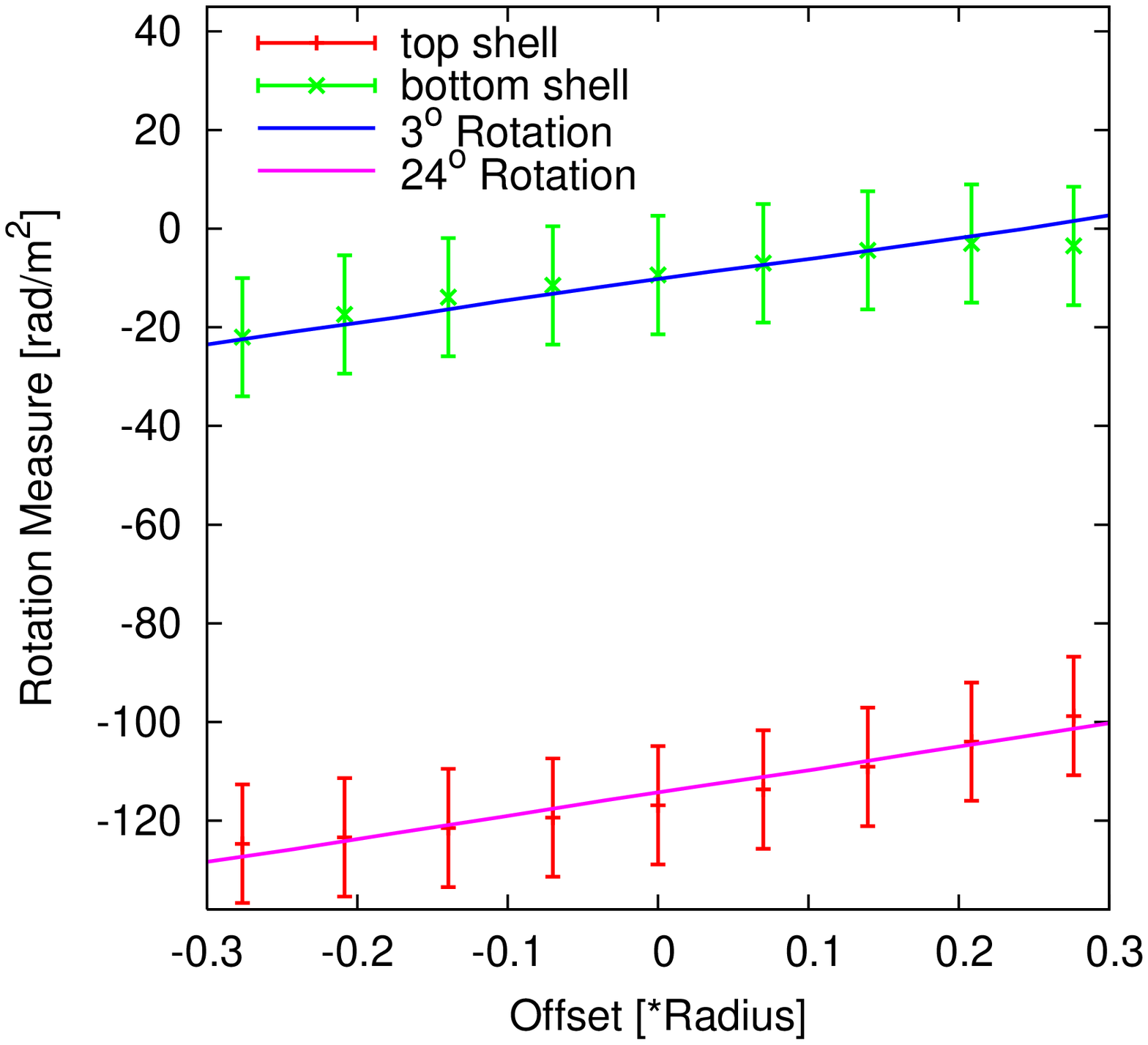}
\end{minipage}
\caption{Left: Polarized intensity map of the SNR DA\,530 at 4.85\,GHz, observed 
with the 100\,m telescope in Effelsberg. Vectors in B-field direction 
are overlaid. Centre: Rotation measure map calculated between 
4.85\,GHz and 10.45\,GHz (100\,m Effelsberg) of SNR DA\,530 with overlaid white
contours indicating polarized intensity. Right: Observed rotation measure 
gradients on the shells of the SNR DA\,530 as a function of distance from the
centre of the shell.}
\label{da530}
\end{figure}

In Fig.~\ref{da530} we display a polarized intensity and rotation measure map of
the SNR DA\,530, which is located high above the Galactic plane at a latitude of
about $+7^\circ$. The overlaid vectors in the polarized intensity image indicate
a tangential magnetic field, which we expected from the simulations for a SNR
with a $\Theta$ of less than $60^\circ$. The rotation measure in the centre is
about 0\,rad/m$^2$, hence we can neglect foreground effects. The RM map is 
dominated by negative rotation measures, but there is a small area of positive RM
to the left of the centre. This indicates that the SNR is expanding into a 
magnetic field, which is pointing away from the front left to the back right.

As can be seen in Fig.~\ref{da530} the RM values on both shells differ 
significantly, however, the gradient seems to be the same. This implies that
the ambient density and magnetic field strength are the same for both shells.
The difference could be either in the foreground RM or the ambient magnetic 
fields for the two shells have a different $\Theta$. To find such a large 
difference in 
rotation measure on such a small scale in the foreground is very unlikely, 
since it would require
either another SNR or an HII region in the foreground to produce such a large
effect both of which would be easily detectable by some other means. The only
possibility left is twisted ambient magnetic field. A simulation for
both shells indicates that the top shell is expanding into a magnetic field
with $\Theta = 24^\circ$ and the bottom one with $\Theta = 3^\circ$. The lower
surface brightness of the top shell supports this finding. The radio surface 
brightness goes down with $\Theta$, because the magnetic field inside the SNR
is more and more along the line of sight (see also Equation~\ref{flux}).

Radio observations of other galaxies show twisted magnetic spurs emerging from
star forming regions (e.g. Review by \cite{beck08}). DA\,530 is located above an 
area of the Milky Way, which is rich in star forming regions, HII regions, and 
SNRs. Is DA\,530 expanding inside these twisted magnetic spurs?

\section{The SNR G182.4+4.3 and the Galactic anti-centre}

\begin{figure}
\begin{center}
\begin{minipage}{5.5cm}
  \includegraphics[bb = 30 30 550 540,height=5.5cm,clip]{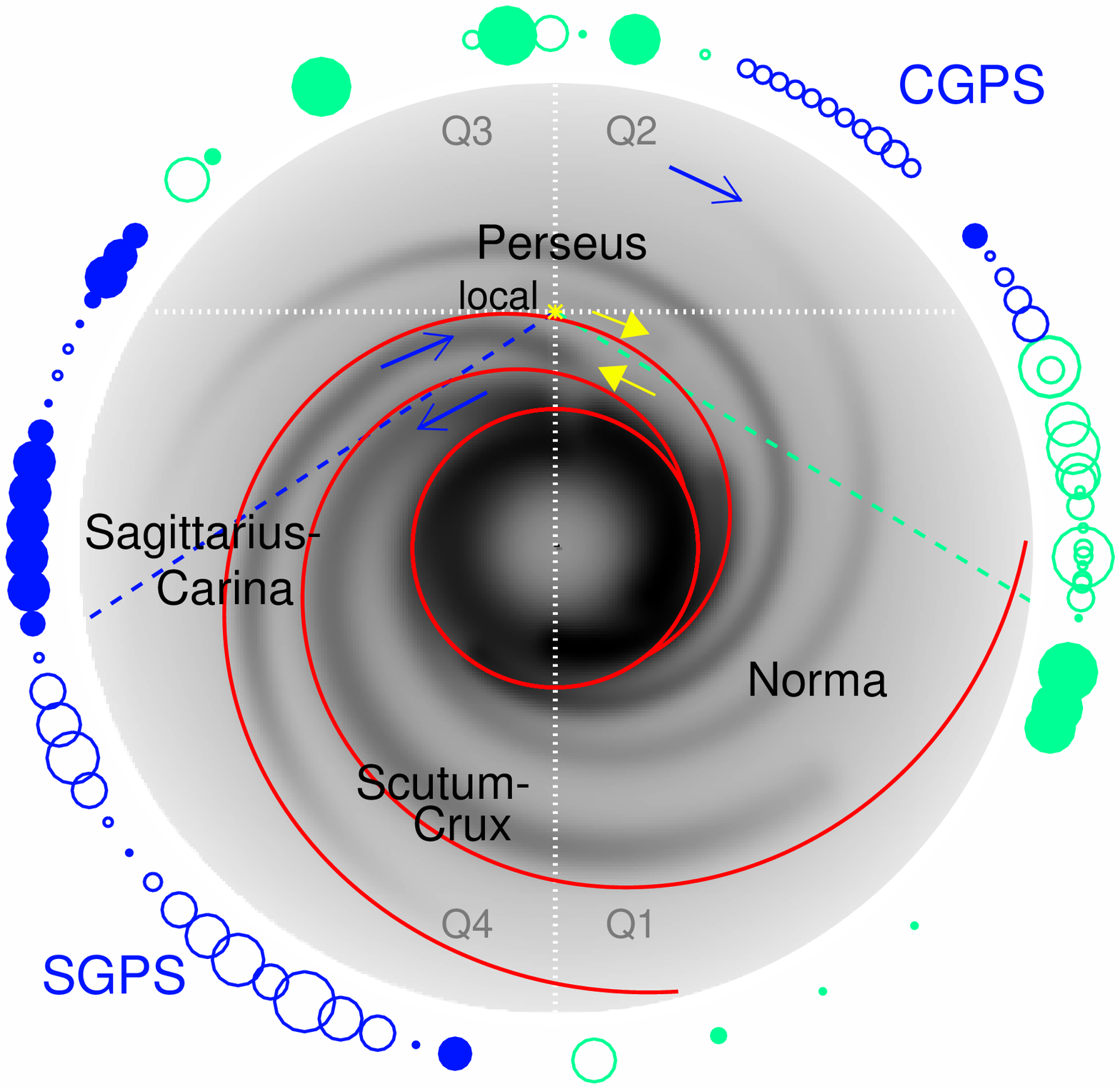}
\end{minipage}
\begin{minipage}{5.5cm}
  \includegraphics[bb = 30 30 550 540,height=5.5cm,clip]{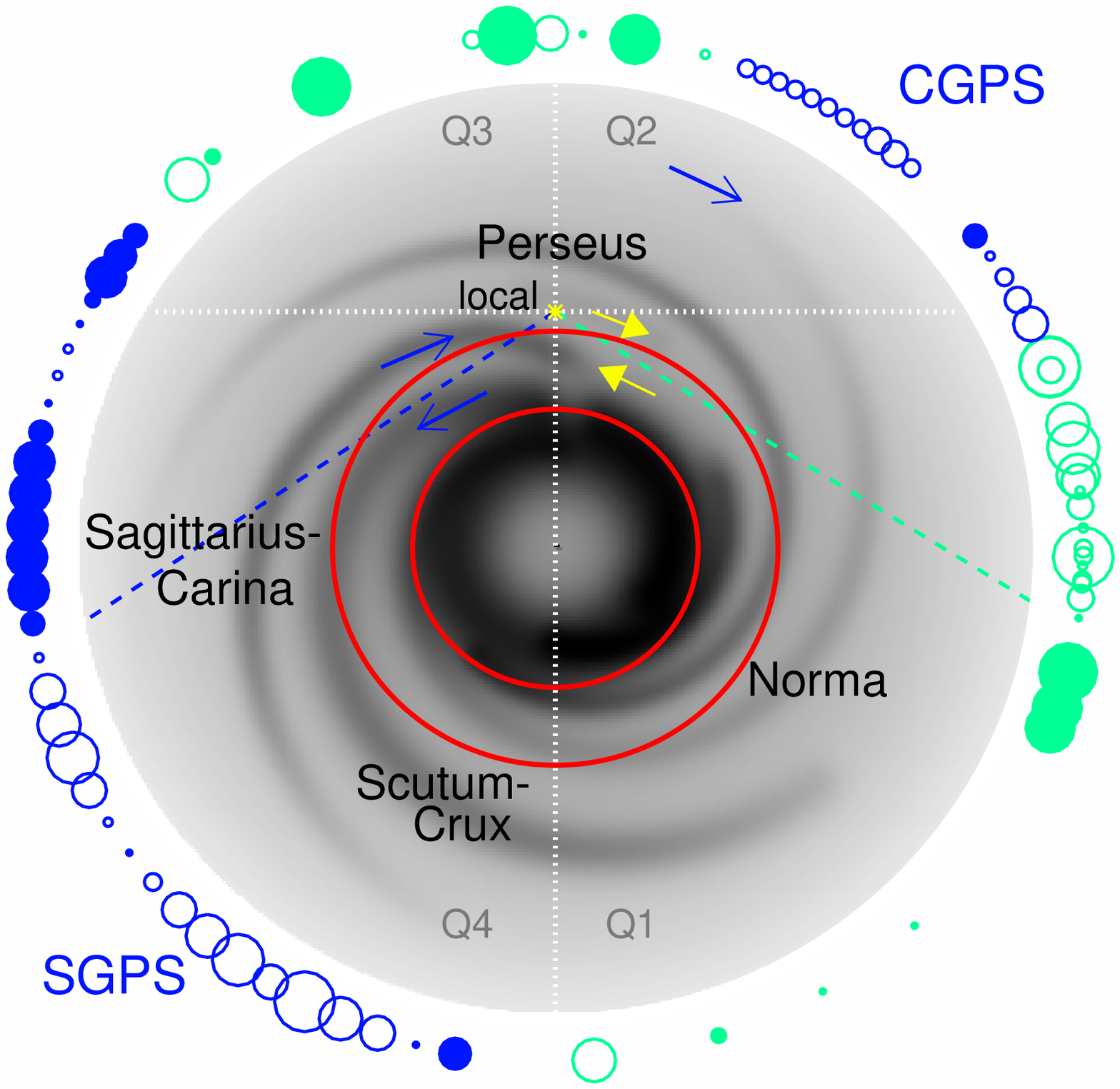}
\end{minipage}
\end{center}
\caption{Extragalactic rotation measures and the large-scale magnetic field 
of the Milky Way - 
a birds-eye-view. Filled symbols indicate positive RM, open circles indicate 
negative RM, and the size of the circles are proportional to the RM. 
The gray-scale is the electron density model of \cite{cl02}.  
The filled arrows indicate magnetic field directions
commonly accepted, while the open arrows indicate magnetic field directions as 
determined from \cite{brown07}.
The solid lines are suggested spiral field lines (left) and circular field
lines (right). The common names of the 
 spiral arms, as well as 
the Galactic quadrants are also labeled.}
\label{mw}
\end{figure}

In terms of the large-scale Galactic magnetic field, the community 
generally agrees that the field is directed clockwise in our local arm, as viewed
from the North Galactic pole, and that there is at least one large-scale 
magnetic field reversal (a region of magnetic shear where the field 
is seen to reverse directions by roughly $180^\circ$) in the inner Galaxy 
between the local and Sagittarius-Carina arms (\cite{sk79},\cite{TN80}),
as indicated by the filled arrows in quadrant 1 (Q1) in 
Fig.~\ref{mw}. In addition to the existence and location of magnetic field 
reversals is the question of field alignment with the optical spiral arms.  It is 
often assumed that the field is closely aligned with the spiral arms. 
\cite{brown07} demonstrated conclusively
that the field in the fourth quadrant (Q4) of the Sagittarius-Carina 
arm is directed clockwise, the same as the local field.  This presents a 
continuity problem if the field does indeed follow the optical arms and is 
counter-clockwise 
in Q1 of the Sagittarius-Carina arm, as shown in Fig.~\ref{mw} (left).
However, if the field is much less inclined than the optical spiral arms, perhaps
even purely azimuthal as suggested by \cite{vallee05}, the continuity problem
is resolved, as shown in Fig.~\ref{mw} (right).
To find additional evidence a good place to look is towards the 
anti-centre region of 
the Milky Way. If the large-scale magnetic field of our Galaxy is azimuthal the
magnetic field component parallel to the line of sight $B_\parallel$ would be 0
towards a Galactic longitude of $180^\circ$ and if the magnetic field follows the
spiral arms we would expect $B_\parallel = 0$ towards Galactic longitude between
$165^\circ$ and $170^\circ$.

\begin{figure}
\begin{minipage}{4cm}
  \includegraphics[bb = 110 105 535 485,height=4cm,clip]{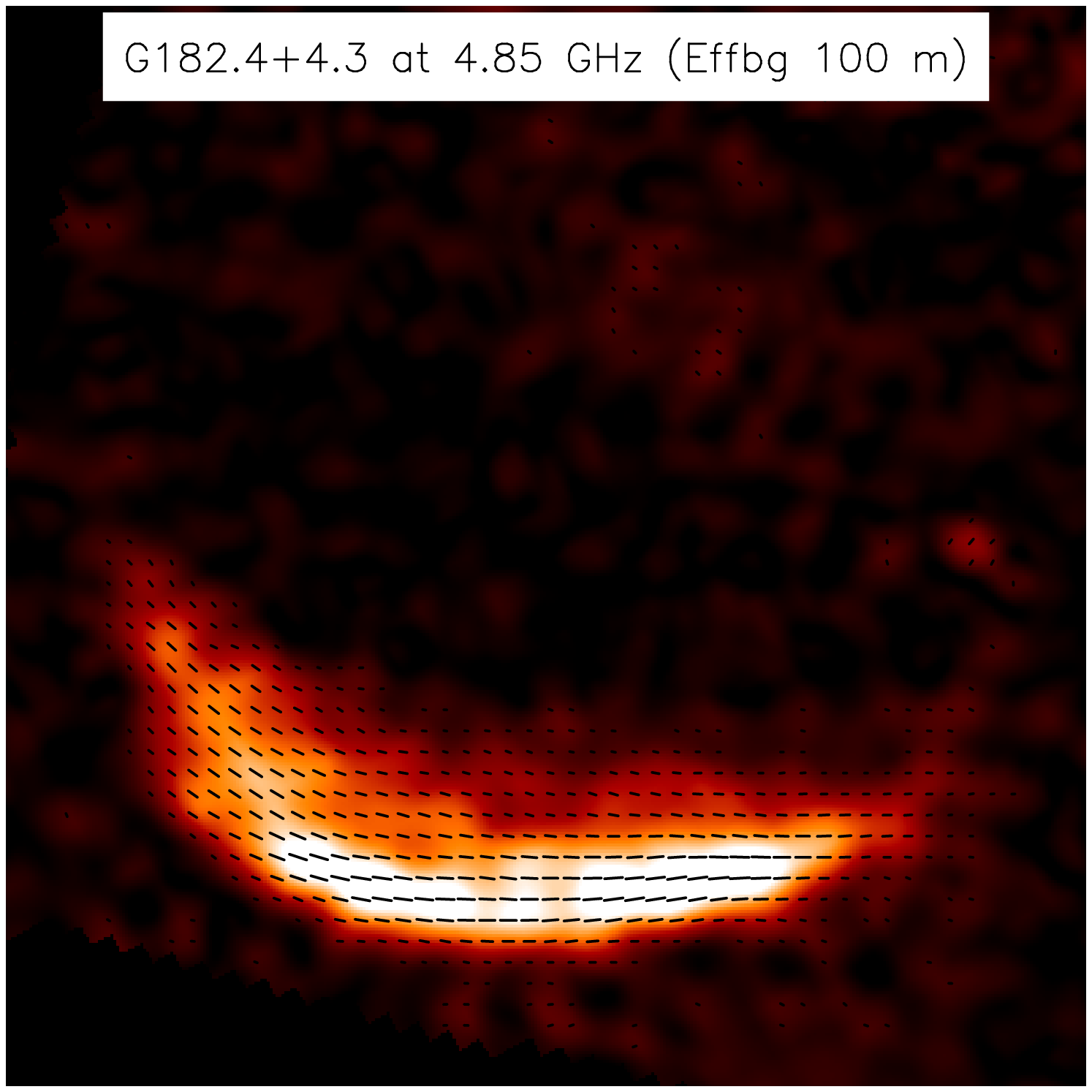}
\end{minipage}
\begin{minipage}{4.8cm}
  \includegraphics[bb = 80 103 525 445,height=4cm,clip]{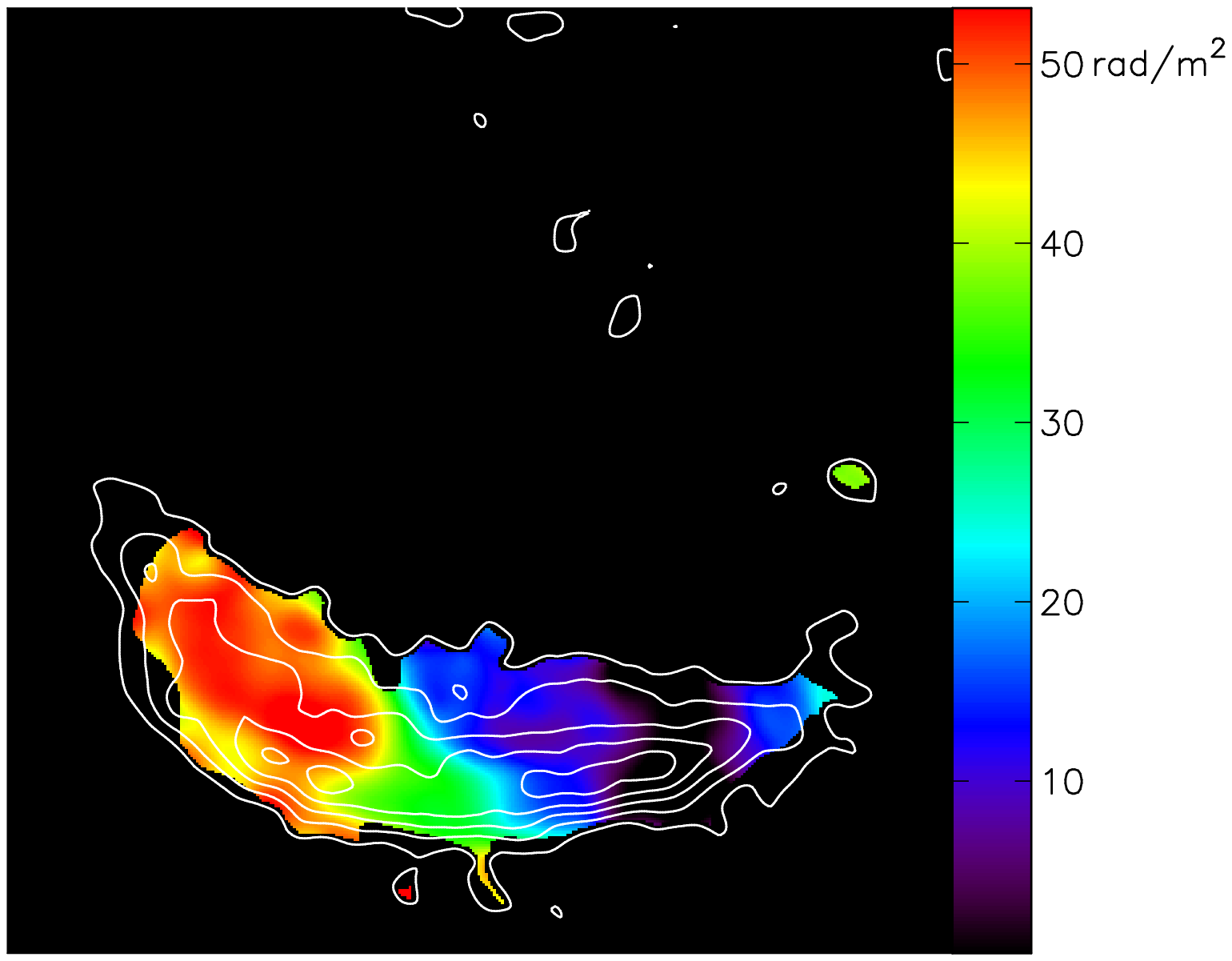}
\end{minipage}
\begin{minipage}{4cm}
  \includegraphics[bb = 55 45 530 500,height=4cm,clip]{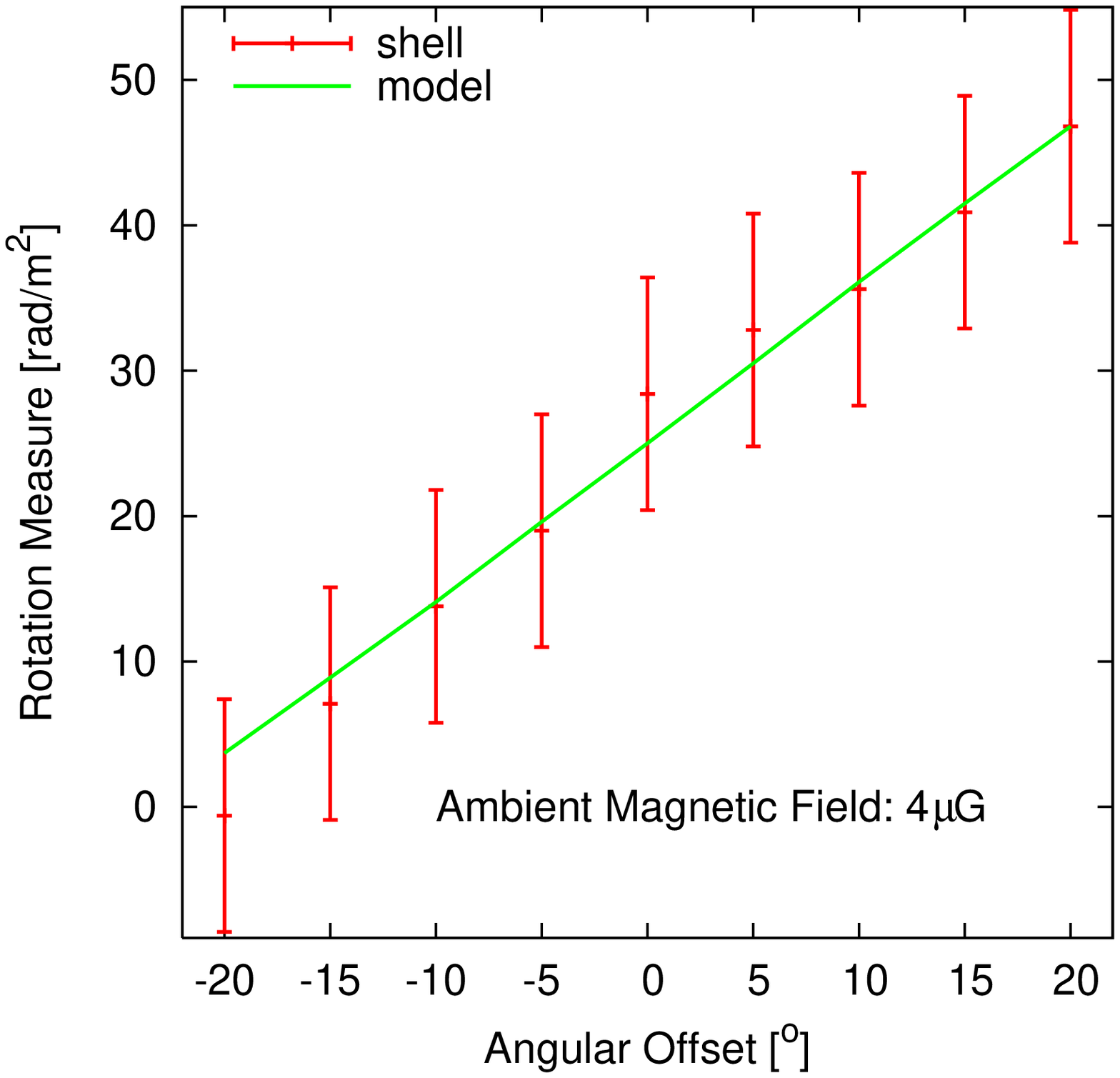}
\end{minipage}
\caption{The same as Fig.~\ref{da530} but for the SNR G182.4+4.3.
The rotation measure map was calculated between 4.85\,GHz (100\,m
Effelsberg) and 1420~MHz taken from the Canadian Galactic Plane Survey
(\cite{tayl03}).}
\label{g182}
\end{figure}

The SNR G182.4+4.3 (Fig.~\ref{g182}) is conveniently located close to the 
anti-centre at a Galactic longitude of $182.4^\circ$. Unfortunately only one shell of this SNR is visible, likely because
the bottom shell is expanding towards the Galactic plane into higher density and
the top shell away from the plane into very low density. However, without emission
from the centre of the SNR we cannot easily determine the foreground $RM$. On
the bottom shell we find a RM gradient which goes from about 0 in the right to
about 50\,rad/m$^2$ on the left (Fig.~\ref{g182}). This implies that the ambient 
magnetic field
is pointing towards us from back left to front right. This is not surprising, 
since the magnetic field in the outer Galaxy should be directed from higher to
lower longitudes and we are here at a longitude which is higher than the 
$B_\parallel = 0$ point for both possible magnetic field configurations. 
Neglecting the foreground contribution for the RM we can determine an upper limit
for $\Theta$ by simulating a SNR for the RM map of G182.4+4.3. The result is
independent of the distance to G182.4+4.3. We derive $\Theta \ge -4^\circ$,
hence the Galactic longitude at which $B_\parallel = 0$ must be larger than
$178^\circ$. If we now assume that the foreground magnetic field has the same
$\Theta$ as the SNR's ambient field and the foreground field is constant at
$4\,\mu$G, which would be a reasonable assumption, we would find 
$\Theta = -2^\circ$. This supports the assumption that the Milky Way has an
azimuthal magnetic field structure.

\begin{acknowledgments}
The Dominion Radio Astrophysical Observatory is a National Facility
operated by the National Research Council Canada. The Canadian Galactic Plane
Survey is a Canadian project with international partners, and is
supported by the Natural Sciences and Engineering Research Council
(NSERC).  This research is based on observations with the 100-m
telescope of the MPIfR at Effelsberg.
\end{acknowledgments}


\begin{thebibliography}{}

\bibitem[Beck (2008)]{beck08}
     {Beck, R.} 2008, \textit{astroph} 0711.4700

\bibitem[Brown \etal\ (2007)]{brown07}
     {Brown, J.C., Haverkorn, M., Gaensler, B.M., Taylor, A.R., Bizunok, N.S., 
     McClure-Griffiths, N.M., Dickey, J.M., \& Green, A.J.} 
     2007, \textit{ApJ} 663, 258

\bibitem[Cordes \& Lazio (2002)]{cl02}
     {Cordes, J.M., \& Lazio, T.J.W.} 2002, \textit{astroph} 0207156

\bibitem[Simard-Normandin \& Kronberg (1979)]{sk79}
     {Simard-Normandin, M., \& Kronberg, P.P.} 1979, \textit{Nature} 279, 115 

\bibitem[Taylor, Gibson, Peracaula, \etal\ (2003)]{tayl03}
     {Taylor, A.R., Gibson, S.J., Peracaula, M., Martin, P.G., Landecker, T.L., 
     Brunt, C., Dewdney, P.E., Dougherty, S.M., Durand, D., Gibson, S.J., 
     Gray, A.D., Higgs, L.A., Kerton, C.R., Knee, L.B.G., Kothes, R., 
     Purton, C.R., Uyan{\i}ker, B., Wallace, B.J., Willis, A.G.} 
     2003, \textit{AJ} 124, 3145

\bibitem[Thomson \& Nelson (1980)]{TN80} 
 {Thomson, R.C., \& Nelson, A.H.} 1980, \textit{MNRAS} 191, 863 

\bibitem[Vall{\'e}e (2005)]{vallee05}
  {Vall{\'e}e, J.P.} 2005, \textit{ApJ} 619, 297
       
\end{thebibliography}
\end{document}